\documentclass[twocolumn,superscriptaddress,showpacs,preprintnumbers,amsmath,amssymb,prl]{revtex4}
\usepackage{graphicx}
\usepackage{dcolumn,xcolor,ulem}
\usepackage{amsmath} 
\usepackage{amssymb}
\usepackage{amsfonts}
\usepackage{bm}
\usepackage[latin1]{inputenc}

\newcommand\gp{\dot\gamma}
\newcommand\gl{\gamma_{\rm loc}}
\newcommand\gpm{\dot\gamma_{\rm min}}
\newcommand\taum{\tau_{\rm min}}

\begin{document}


 \title{Creep and fracture of a protein gel under stress}

\author{Mathieu Leocmach}
\email{mathieu.leocmach@ens-lyon.fr}
\affiliation{Universit\'e de Lyon, Laboratoire de Physique, \'Ecole Normale Sup\'erieure de Lyon, CNRS UMR 5672, 46 All\'ee d'Italie, 69364 Lyon cedex 07, France}
\author{Christophe Perge}
\affiliation{Universit\'e de Lyon, Laboratoire de Physique, \'Ecole Normale Sup\'erieure de Lyon, CNRS UMR 5672, 46 All\'ee d'Italie, 69364 Lyon cedex 07, France}
\author{Thibaut Divoux}
\affiliation{Centre de Recherche Paul Pascal, CNRS UPR 8641 - 115 avenue Schweitzer, 33600 Pessac, France}
\author{S\'ebastien Manneville}
\affiliation{Universit\'e de Lyon, Laboratoire de Physique, \'Ecole Normale Sup\'erieure de Lyon, CNRS UMR 5672, 46 All\'ee d'Italie, 69364 Lyon cedex 07, France}

\date{\today}

\begin{abstract}
Biomaterials such as protein or polysaccharide gels are known to behave qualitatively as soft solids and to rupture under an external load. Combining optical and ultrasonic imaging to shear rheology we show that the failure scenario of a protein gel is reminiscent of brittle solids: after a primary creep regime characterized by a power-law behavior which exponent is fully accounted for by linear viscoelasticity, fractures nucleate and grow logarithmically perpendicularly to shear, up to the sudden rupture of the gel. A single equation accounting for those two successive processes nicely captures the full rheological response. The failure time follows a decreasing power law with the applied shear stress, similar to the Basquin law of fatigue for solids. These results are in excellent agreement with recent fiber-bundle models that include damage accumulation on elastic fibers and exemplify protein gels as model, brittle-like soft solids.
\end{abstract}

\pacs{82.35.Pq, 47.57.Qk, 83.60.-a, 83.80.Kn}
\maketitle

Biogels formed through the self-association of polysaccharide coils, collagen, actin filaments or attractive globular proteins play a major role in biochemistry and microbiology \cite{Viovy:2000}, biological networks and cell mechanics \cite{Stricker:2010} as well as in food science \cite{Mezzenga:2005}. These biomaterials all behave as elastic solids under small deformations but display remarkable nonlinear behavior generally featuring stress- or strain-stiffening \cite{Gardel:2004} and fractures prior to irreversible rupture \cite{Bonn:1998,Baumberger:2006}. Irreversibility stems from the existence of an external control parameter, e.g. temperature or pH in the case of thermoreversible or acid-induced gels respectively. This makes such biogels fundamentally different from other soft glassy materials such as emulsions, colloidal gels and glasses that can be rejuvenated by shear \cite{Cloitre:2000,Caton:2008,Divoux:2011,Siebenburger:2012} or transient networks where fractures spontaneously heal \cite{Tabuteau:2009,Skrzeszewska:2010}. So far, huge effort has been devoted to the design of protein gels with specific properties and textures at rest \cite{Dickinson:2006,Gibaud:2012}. However, their mechanical behavior deep into the nonlinear regime has only been partially addressed \cite{vanVliet:1995,Pouzot:2006} and several fundamental issues remain unexplored such as the spatially resolved rupture scenario or the physical relevance of the analogy with brittle failure in hard solids.

\begin{figure}
\centering
\includegraphics[width=7cm]{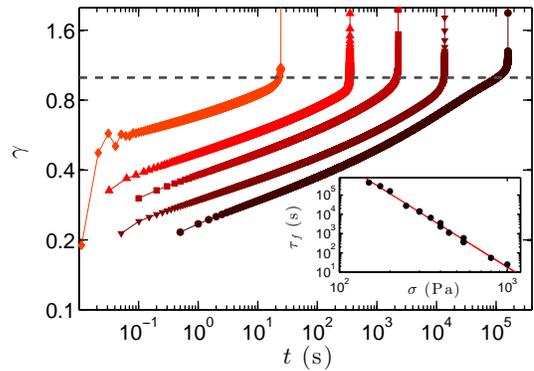}
\caption{(color online) Strain response $\gamma(t)$ in a 4\%~wt. casein gel acidified with 1\%~wt. GDL for an imposed shear stress $\sigma=200$ (\textcolor{red!25!black}{$\bullet$}), 300 (\textcolor{red!50!black}{$\blacktriangledown$}), 400 (\textcolor{red!75!black}{$\blacksquare$}), 550 (\textcolor{red}{$\blacktriangle$}) and 1000~Pa (\textcolor{orange!50!red}{$\blacklozenge$}) from right to left. The gap width is 1~mm. Gray dashes show $\gamma=1$. Inset: failure time $\tau_f$ vs $\sigma$. The red line is the best power-law fit $\tau_f=A\sigma^{-\beta}$ with $\beta=5.45\pm 0.05$ and $A=(4.2\pm 0.1)\,10^{17}$~s.Pa$^\beta$.
\label{fig1}}
\end{figure} 

In this Letter we report on stress-induced fracture in protein gels by means of creep experiments coupled to optical and ultrasonic imaging. Gels formed by slow acidification of a sodium caseinate solution display fractures under large strain at fixed low pH values \cite{vanVliet:1995,Lucey:1998}, which makes them perfect candidates to quantify the rupture of soft solids and tackle the above-mentioned issues. We demonstrate that under an external load, these casein gels display brittle-like failure that results from two successive physical processes: (i) a primary creep regime where dissipation is dominated by viscous flow through the gel matrix without any detectable macroscopic strain localization and (ii) the irreversible nucleation and growth of fractures leading to gel failure. Our results are in full agreement with the predictions of some recent fiber-bundle models and hint to universal features of failure common to both soft and hard solids.

\begin{figure}
\centering
\includegraphics[width=7cm,clip]{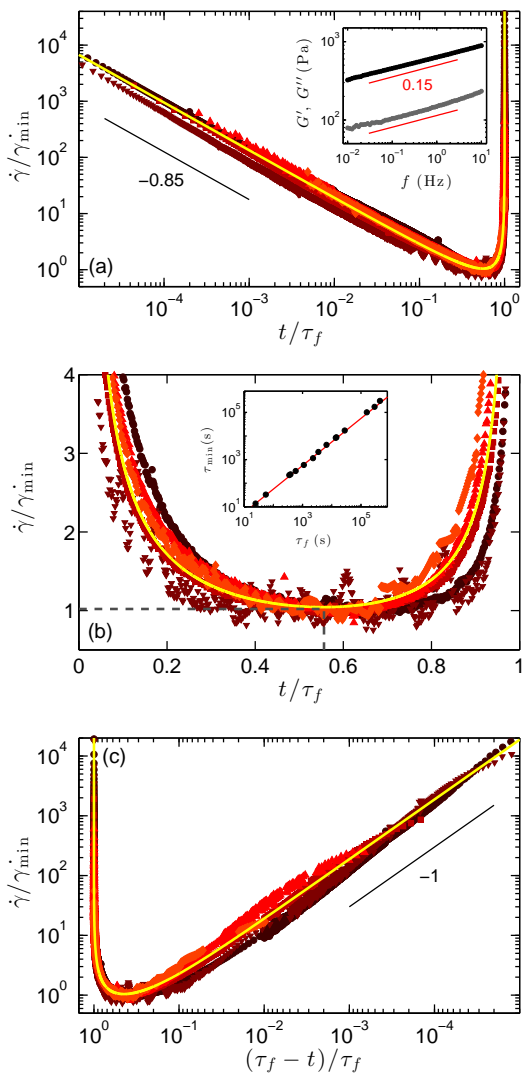}
\caption{(color online) Normalized shear rate responses $\gp(t)/\gpm$ corresponding to the data of Fig.~\ref{fig1} and plotted so as to emphasize the three successive regimes. $\gpm$ is the minimum shear rate reached at $\taum$ (see text and Suppl. Fig.~2). The yellow line shows the master curve inferred from fitting $\gp(t)$ by Eq.~(\ref{e.fit}) with $\alpha=0.85$, leading to $\lambda=0.378\pm 0.002$ and $\mu=0.187\pm 0.002$. (a)~Primary creep: $\gp(t)/\gpm$ vs $t/\tau_f$ in logarithmic scales. Inset: Linear viscoelastic moduli $G'$ (top) and $G''$ (bottom) as a function of frequency $f$ for a strain amplitude of 0.1\%. Red lines are power laws $G'\sim G''\sim f^{0.15}$. (b)~Secondary creep: $\gp(t)/\gpm$ vs $t/\tau_f$ in linear scales. Gray dashes show the minimum of Eq.~(\ref{e.fit}) reached at $\taum=0.556\tau_f$. Inset: $\taum$ vs $\tau_f$. The red line is $\taum=0.56\tau_f$. (c)~Tertiary creep: $\gp(t)/\gpm$ vs $(\tau_f-t)/\tau_f$ in logarithmic scales with a reversed horizontal axis.\label{fig2}}
\end{figure} 

Gels are prepared by dissolving sodium caseinate powder (Firmenich) at 4\% wt. in deionized water under gentle mixing at 35$^{\circ}$C and 500~rpm. To induce gelation, 1\% wt. glucono-$\delta$-lactone (GDL) in powder (Firmenich) is dissolved in the solution and its hydrolysis progressively lowers the pH over the course of 8~hours [see Fig.~1(a) in the Supplemental Material, including Ref.~\cite{Dickinson:2002supp}]. While still liquid, the solution is poured into the gap of a polished Plexiglas concentric cylinder (or Taylor-Couette, TC) shear cell immersed into a temperature-controlled water tank at 25.0$\pm$0.1$^{\circ}$C \cite{note}. Rheological data are recorded during gel formation by a stress-controlled rheometer through small amplitude oscillatory shear at frequency $f=1$~Hz (Fig.~1 in the Supplemental Material). Gelation is complete when the elastic ($G'$) and viscous ($G''$) moduli reach a plateau with $G'\gg G''$. A constant stress $\sigma$ is then applied to the sample from time $t=0$ and the subsequent strain response $\gamma(t)$ is monitored. Images of the gel are recorded simultaneously to the rheology (Logitech Webcam Pro 9000). The local velocity and strain fields can also be imaged in the gradient--vorticity plane $(r,z)$ simultaneously to rheology by a custom-made ultrasonic scanner detailed in \cite{Gallot:2013}. In this case, prior to acidification, the sodium caseinate solution is seeded with acoustic tracers here 3\% wt. polyamide spheres (Orgasol 2002 ES3 NAT 3, Arkema, diameter 30~$\mu$m, density 1.02) that do not modify the final gel properties [Fig.~1(b) in the Supplemental Material]. The radial position $r$ is measured from the inner rotating cylinder and the vertical axis $z$ points downwards with the origin about 15~mm from the top of the TC cell. Failure being irreversible, each creep experiment requires to prepare a fresh sample in-situ.  

\begin{figure*}[t]
\centering
\includegraphics[width=17.5cm,clip]{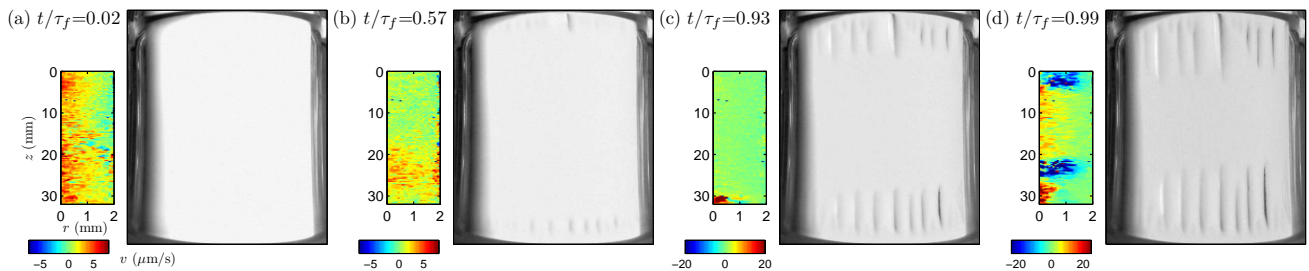}
\caption{(color online) Ultrasonic images (left) and direct visualization of the TC cell (right) recorded simultaneously at various times in the primary [(a)~$t/\tau_f=0.02$], secondary [(b)~$t/\tau_f=0.57\simeq\taum/\tau_f$] and tertiary  [(c) $t/\tau_f=0.93$ and (d) 0.99] creep regimes. Ultrasonic images are azimuthal velocity maps $v(r,z,t)$ computed by averaging over 4~s and coded using the linear color levels shown below the images. Their position relative to the picture of the cell reflects the actual arrangement of the ultrasonic probe along the vertical direction $z$. The gap width is 2~mm and $r$ is the radial distance to the inner rotating cylinder. Experiment performed on a 4\%~wt. casein gel seeded with 3\%~wt. polyamide spheres and acidified with 1\%~wt. GDL under $\sigma=300$~Pa. See also Movie~2 in the Supplemental Material.
\label{fig3}}
\end{figure*}

\begin{figure*}
\centering
\includegraphics[width=17.5cm,clip]{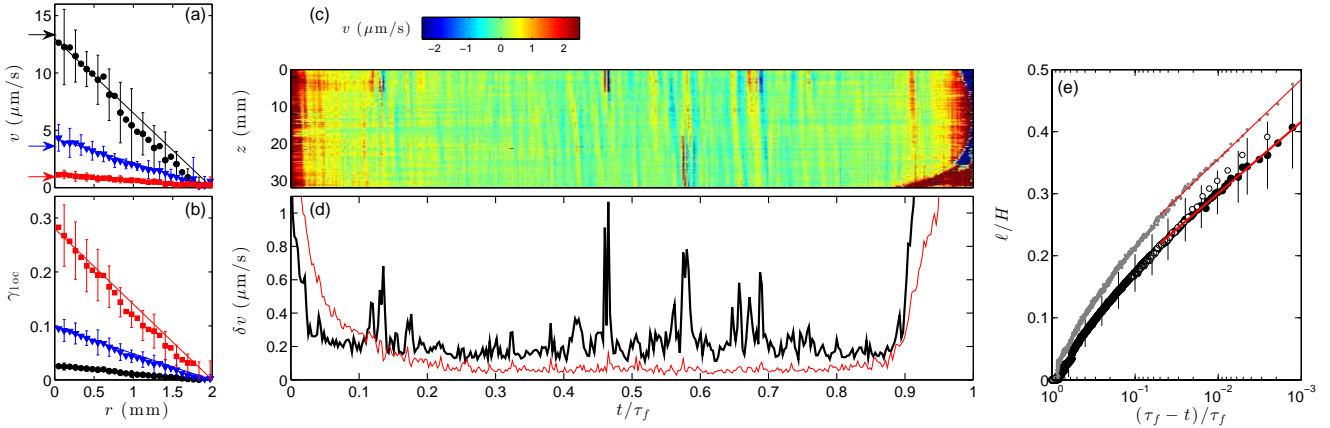}
\caption{(color online) (a)~Local velocity $\langle v(r,z,t)\rangle_z$ and (b)~local strain field $\langle\gl(r,z,t)\rangle_z$ averaged over the vertical direction $z$ at various times during primary creep: $t/\tau_f=1.9\,10^{-3}$ ($\bullet$), $1.7\,10^{-2}$ (\textcolor{blue}{$\blacktriangledown$}) and 0.15 (\textcolor{red}{$\blacksquare$}). Solid lines are linear profiles. The arrows in (a) indicate the velocity of the inner cylinder inferred from the current shear rate. (c)~Spatiotemporal diagram of the local velocity $\langle v(r,z,t)\rangle_r$ averaged over the radial direction $r$ and plotted in linear color levels as a function of $z$ and $t/\tau_f$. (d)~Standard deviation $\delta_z v(t)$ of $\langle v(r,z,t)\rangle_r$ taken over the vertical direction $z$ (thick black line) together with corresponding standard deviation $\delta_r v(t)$ computed over the radial direction $r$ on the $z$-average $\langle v(r,z,t)\rangle_z$ (thin red line). (e)~Fracture length $\ell(t)$ vs $(\tau_f-t)/\tau_f$ as inferred from direct visualization ($\bullet$, average over 6 different fractures, error bars show the standard deviation) and from ultrasonic imaging ($\circ$) and normalized by the height $H$ of the TC cell. Gray dots show the visualization data for the longest fracture which leads to the failure of the sample at $\tau_f$. Red lines are the best fits $\ell(t)=a+b\log(1-t/\tau_f)$ to the visualization data. Same experiment as in Fig.~\ref{fig3} and Supplemental Movie~2.
\label{fig4}}
\end{figure*}

Under a constant applied shear stress $\sigma$, the global strain $\gamma(t)$ displays a robust time dependence (Fig.~\ref{fig1}): $\gamma(t)$ slowly grows with time up to $\gamma \sim 1$ then accelerates until the gel fails at a well-defined time $\tau_f$. These three successive steps are better highlighted in Fig.~\ref{fig2} by focusing on the global shear rate $\gp(t)$. Figures~\ref{fig3} and \ref{fig4} gather the results from {\it local} measurements and are discussed below together with each of the successive regimes inferred from {\it global} data. Note that similar failure dynamics are observed for other casein and GDL concentrations, for different gap widths of the TC cell and in a different geometry namely a plate-plate cell (see Movie~1 in the Supplemental Material).

As seen in the inset of Fig.~\ref{fig1}, the failure time $\tau_f$ sharply decreases as a power law of $\sigma$ with an exponent $\beta\simeq 5.5$. $\tau_f$ further allows us to rescale all the shear rate data $\gp(t)$ onto the single master curve of Fig.~\ref{fig2}(a) by plotting $\gp/\gp_{\rm min}$ vs $t/\tau_f$, where $\gp_{\rm min}$ is the minimum shear rate reached at a time $\tau_{\rm min}$ [see also Fig.~2 in the Supplemental Material for the original $\gp(t)$ data]. On about four decades for $t\lesssim 0.1\tau_f$, the shear rate decreases as a power law $\gp(t)\sim t^{-\alpha}$ with $\alpha=0.85\pm 0.04$. This is strongly reminiscent of the {\it primary creep} observed in solids and referred to as Andrade creep \cite{Andrade:1910,Miguel:2002,Nechad:2005,Rosti:2010}. Interestingly, here, the exponent can be inferred from linear viscoelasticity. Indeed casein gels display a power-law rheology $G'(f)\sim G''(f)\sim f^{0.15}$ [inset of Fig.~\ref{fig2}(a)], which corresponds to a compliance $J(t)\equiv\gamma(t)/\sigma\sim t^{0.15}$ in the linear deformation regime \cite{Tschoegl:1989}. This nicely corresponds to the observed $\gp(t)\sim t^{-0.85}$. Moreover ultrasonic imaging reveals velocity and strain fields averaged over the vertical direction $z$ that linearly decrease with the radial position $r$ within the gap [Fig.~\ref{fig4}(a-b)] with no slippage at the Plexiglas walls [see arrows in Fig.~\ref{fig4}(a)]. Together with direct visualization [Fig.~\ref{fig3}(a)], these local measurements demonstrate that during primary creep there is no macroscopic strain localization or fracture, although we cannot rule out rearrangements below the available spatial ($\sim10~\mu$m) and/or temporal ($\sim1$~s) resolutions due to limited signal-to-noise ratio at very low shear rates [Fig.~\ref{fig3}(a, left)].

For $0.1\lesssim t/\tau_f\lesssim 0.9$, $\gp(t)$ departs from power-law behavior and goes through a minimum value at time $\tau_{\rm min}= (0.56\pm0.04)\tau_f$ independently of the applied stress [Fig.~\ref{fig2}(b) and inset] as similarly reported for metals \cite{Sundararajan:1989}, solid composite materials \cite{Nechad:2005} and fiber-bundle models (FBMs) \cite{Kovacs:2008,Jagla:2011}. This linearity between $\tau_{\rm min}$ and $\tau_f$, also known as the Monkman-Grant relation \cite{Monkman:1956}, allows one to ``predict'' the failure time from the intermediate-time response. During this {\it secondary creep} regime, regularly-spaced cracks nucleate from the top and bottom edges of the cell and start growing perpendicular to the applied stress [Fig.~\ref{fig3}(b), see also Movie~2 in the Supplemental Material]. These macroscopic fractures are invaded with water expelled from the surrounding gel matrix. Although fractures have not yet entered the ultrasonic region of interest, velocity maps become heterogeneous along the $z$ direction [Fig.~\ref{fig3}(b, left)] and display intermittent fluctuations [Fig~\ref{fig4}(c,d)]. The level of these fluctuations, which can also be seen on the global response for the same applied stress [\textcolor{red!50!black}{$\blacktriangledown$} in Fig.~\ref{fig2}(b)], is poorly reproducible and stress-dependent. Such intermittency may arise from crack growth outside the region of interest which induces long-range elastic deformation propagating across the sample mainly along the vertical direction [see the lower level of fluctuations along the radial direction in Fig~\ref{fig4}(d)].

In the {\it tertiary creep} regime, for $t\gtrsim 0.9\tau_f$, the shear rate increases by more than four orders of magnitude and diverges as $(\tau_f-t)^{-1}$ [Fig.~\ref{fig2}(c)]. This finite-time singularity corresponds to the final growth of the fractures along the vorticity direction $z$ as they accelerate and eventually meet in the middle of the cell at time $\tau_f$ [Fig.~\ref{fig3}(c-d)]. Ultrasonic velocity maps directly correlate with the crack growth and reveal the complex structure of the displacement field at the tip and around the fracture. In particular, Fig.~\ref{fig3}(d, left) suggests that the fracture is initiated at the inner cylinder at $r=0$ and the presence of large positive and negative velocities in the vicinity of the crack tip is indicative of strong compression and recoil of the gel matrix. Finally, the fracture length $\ell(t)$ is observed to grow logarithmically with $(\tau_f-t)$ upon approaching $\tau_f$ [Fig~\ref{fig4}(d)]. In other words one has ${\rm d}\ell/{\rm d}t\sim\gp(t)\sim (\tau_f-t)^{-1}$, which indicates that the global shear rate is linked to fracture-induced displacements.

Finally, we emphasize that the Andrade-like creep and the crack growth are two physical processes that effectively superimpose to yield the global rheological response. Indeed, as seen from the yellow line in Fig.~\ref{fig2}, the master curve $\gp(t)/\gp_{\rm min}$ vs $t/\tau_f$ is perfectly fitted by:
\begin{equation}
\frac{\gp(t)}{\gp_{\rm min}}=\lambda \left(\frac{t}{\tau_f}\right)^{-\alpha}+ \frac{\mu}{1-t/\tau_f}\,,
\label{e.fit}
\end{equation}
with only two adjustable parameters $\lambda$ and $\mu$ once $\alpha=0.85$ is fixed. The remarkable collapse of the whole data set to such a simple equation allows us to interpret the secondary creep regime as a mere crossover from creep to crack growth.

Let us now summarize and discuss the most prominent results of this Letter. First we have shown that casein gels display a remarkable failure scenario similar to that of brittle solids and characterized by the same three successive creep regimes. Here, the power-law exponent of the primary creep [$\gp(t)\sim t^{-\alpha}$ with $\alpha=0.85\pm 0.04$] is fully accounted for by linear viscoelasticity. Such a link between creep and viscoelasticity is shared by other biopolymer gels with power-law rheology \cite{Gobeaux:2010,Jaishankar:2013} as well as hard-sphere-like colloidal glasses \cite{Siebenburger:2012}. To better understand this link, additional creep and recovery tests were performed within the primary regime (see Figs.~3 and 4 in the Supplemental Material, including Ref.~\cite{Benmouffok-Benbelkacem:2010}). Successive loading/unloading sequences on a fresh sample show that (i)~the strain is not fully recovered by a few percents, (ii)~the viscoelastic properties of the material are unaltered and (iii)~strain responses are indistinguishable from one test to another. This strongly suggests that the small irreversibility observed in the primary regime mainly stems from viscous dissipation (which accounts for the nonzero viscous modulus $G''$) due to the solvent flow within the gel fibrous matrix and involves only a minute amount of local damage of the gel network. Protein gels thus appear to experience a {\it kinetically} reversible primary creep, i.e. dominated by viscous dissipation rather than by plasticity.

The logarithmic fracture growth in the tertiary creep regime constitutes our second important result. Such an evolution is also commonly reported in disordered solid materials displaying brittle rupture and interpreted in the framework of Griffith-like models based on global or local energy barriers \cite{Vanel:2009}. However these approaches all predict exponential scalings for $\tau_f(\sigma)$ while our data are best fitted by the decreasing power law $\tau_f \sim \sigma^{-\beta}$ with $\beta=5.45\pm 0.05$ [inset of Fig.~\ref{fig1}]. This last key result suggests that thermally activated crack growth does not play any prominent role. Rather the power-law scaling is strikingly reminiscent of the Basquin law of fatigue found for a variety of heterogeneous or cellular materials under cyclic deformation \cite{Kun:2007,Basquin:1910,Kohout:2000}. Basquin law has also been recently predicted for creep experiments by FBMs that combine elastic fibers with a local yield strain and take into account damage accumulation \cite{Kun:2007,Halasz:2012}. Interestingly assuming damage accumulation to be proportional to $\sigma^\gamma$ directly leads to Basquin law with $\beta=\gamma$ and large values of $\gamma$, typically larger than 5 as found here for $\beta$, lead to macroscopic cracks due to the simultaneous rupture of a large number of fibers \cite{Halasz:2012}. More generally FBMs under elongational load predict three successive creep regimes exactly alike Fig.~\ref{fig2} with similar proportionality between $\tau_{\rm min}$ and $\tau_f$ and finite-time singularity \cite{Nechad:2005,Jagla:2011}. This highlights the relevance of FBMs in the context of creep in protein gels, whose microstructure indeed appears to be formed of strands \cite{Kalab:1983,Roefs:1990}, but also urges to study FBMs in shear geometries to check whether they would be able to predict fracture growth as observed here.

To conclude, the present time- and space-resolved study exemplifies protein gels as model, brittle-like soft solids. Prompted by the remarkable simplicity of Eq.~(\ref{e.fit}) that encompasses Andrade-like creep, the Monkman-Grant relation and finite-time singularity within a single equation, future modeling will undoubtedly focus on the microscopic ingredients needed to predict quantitatively the Monkman-Grant prefactor and the Basquin exponent $\beta$. The next experimental step consists in a statistical study of the fluctuations associated to crack nucleation and growth as well as a systematic investigation of other systems in order to check for universality in the irreversible creep rupture of soft solids in general and of biogels such as actin, alginate or agar gels in particular. Such a study is expected to have important implications in understanding the behavior of biomaterials under extreme stress conditions.

\begin{acknowledgments}
The authors thank M. Alava, T.~Gibaud, S.~Lindstr\"om, G.~H.~McKinley and N.~Taberlet for fruitful discussions, M.~Kal\'ab for kindly providing us with unobtainable references, and A.~Parker at Firmenich for precious advice and for providing the casein and GDL. This work was funded by the Institut Universitaire de France and by the European Research Council under the European Union's Seventh Framework Programme (FP7/2007-2013) / ERC grant agreement No.~258803. 
\end{acknowledgments}

\end{document}